\documentclass{iaus}
\usepackage{graphicx}

\title[Tests of Planet Formation Models] 
{Observational Tests of Planet Formation Models}

\author[A. Sozzetti {\em et al.}]   
{A. Sozzetti$^{1,2}$, D. W. Latham$^{1}$, G. Torres$^{1}$, B. W. Carney$^{3}$, J. B. Laird$^{4}$, 
R. P. Stefanik$^{1}$, A. P. Boss$^{5}$, D. Charbonneau$^{1}$, F. T. O'Donovan$^{6}$, 
M. J. Holman$^{1}$, \and J. N. Winn$^{7}$}

\affiliation{$^1$Harvard-Smithsonian Center for Astrophysics, Cambridge, MA 02138, USA\\ 
email: {\tt asozzett@cfa.harvard.edu}\\[\affilskip]
$^2$INAF- Astronomical Observatory of Torino, I-10025 Pino Torinese (TO), Italy  \\[\affilskip]
$^3$University of North Carolina at Chapel Hill, Chapel Hill, NC 27599, USA  \\[\affilskip]
$^4$Bowling Green State University, Bowling Green, OH 43403, USA \\[\affilskip]
$^5$Carnegie Institution of Washington, Washington DC 20015, USA \\[\affilskip]
$^6$California Institute of Technology, Pasadena, CA 91125, USA\\[\affilskip]
$^7$Massachusetts Institute of Technology, Cambridge, MA 02139, USA
}

\pubyear{2008}
\volume{249}  
\pagerange{???--???}
\jname{Exoplanets: Detection, Formation, and Dynamics}
\editors{A.C. Editor, B.D. Editor \& C.E. Editor, eds.}
\begin{document}

\maketitle

\begin{abstract}
We summarize the results of two experiments to address important issues 
related to the correlation between planet frequencies and properties and 
the metallicity of the hosts. Our results can usefully inform formation, 
structural, and evolutionary models of gas giant planets.
\end{abstract}

\firstsection 
\section{Introduction}

The planet-metallicity connection is one of the most important aspects of the 
close relationship between characteristics and frequencies of planetary systems 
and the physical properties of the host stars which have been unveiled by the 
present sample of over 200 extrasolar planets. In particular, the likelihood of 
finding a planet around a given star rises sharply with stellar metallicity 
(e.g., Santos et al. 2004; Fischer \& Valenti 2005). 
Furthermore, a correlation may also exist between estimated inner core masses 
of transiting giant planets and the hosts' metal content (Guillot et al. 2006; Burrows 
et al. 2007). In both cases, the 
evidence collected so far appears to strongly support the more widely accepted mechanism 
of giant planet formation by core accretion (e.g., Pollack et al. 1996), 
as opposed to the alternative formation mode by disk instability (e.g., Boss 1997). 
However, the relatively small numbers of metal-poor 
stars screened for planets so far, and the large uncertainties often present 
in the determination of both planet and stellar properties in transiting systems 
prevent one from drawing conclusions.
We describe two experiments designed to put the observed trends on 
firmer observational grounds, thus ultimately helping to discriminate between 
proposed planet formation models. The first is a Doppler survey for giant planets 
orbiting within 2 AU of a well-defined sample of 200 
field metal-poor dwarfs. Our data can help to gauge the behavior of 
planet frequency in the metal-poor regime. The second consists of a novel method 
for improving on the knowledge of stellar and planetary parameters of transiting 
systems through a careful analysis of spectro-photometric measurements. With this 
approach, structural and evolutionary models of irradiated planets can be better 
informed, allowing for refined estimates of the heavy-element content of transiting 
planets and for improved understanding ot the core mass - stellar metallicity correlation.

\section{Testing the $f_p$-[Fe/H] correlation}

Based on an analysis of over 3 yr of precision radial velocity measurements 
of $\sim 200$ metal-poor stars observed with HIRES on the Keck 1 telescope (Sozzetti et al. 2006), 
we have identified several long-term, low-amplitude radial-velocity variables, 
which we are following up with direct imaging techniques at infrared wavelengths. 
We have placed upper limits on the detectable companion mass as a function of orbital 
period. Our survey would have detected, with a $99.9$ confidence level, over 
$95\%$ of all companions on low-eccentricity orbits with velocity semi-amplitude 
$K \gtrsim 100$ m s$^{-1}$, or $M_p\sin i\gtrsim 4.1\,M_\mathrm{J}(P/yr)^{(1/3)}$, 
for orbital periods $P\lesssim 3$ yr. 
None of the stars in our sample exhibits radial-velocity variations compatible 
with the presence of Jovian planets with periods shorter than the survey duration. 
The resulting average frequency of gas giants orbiting metal-poor dwarfs with 
$-2.0\lesssim$[Fe/H]$\lesssim -0.6$ is $f_p<0.67\%$. 
We examine the implications of this null result in the context of the observed 
correlation between the rate of occurrence of giant planets and the metallicity 
of their main-sequence solar-type stellar hosts. By combining our dataset with the 
Fischer \& Valenti (2005) 
uniform sample, we confirm that the likelihood of a star to harbor a planet more massive 
than Jupiter within 2 AU can be expressed as a quadratic function of the host's metallicity. 
However, the data for stars with $-1.0\lesssim$[Fe/H]$\lesssim 0.0$ are compatible, 
in a statistical sense, with a constant occurrence rate $f_p\simeq 1\%$. 

\section{Testing the $M_c$-[Fe/H] correlation}

Using high-resolution Keck spectra, we derive improved atmospheric parameters for the parent stars of 
the recently discovered transiting planets TrES-3 (O'Donovan e al. 2006) and 
TrES-4 (Mandushev et al. 2007). TrES-3 is a mildly metal-deficient 
G-dwarf with [Fe/H] $= -0.19\pm 0.08$, while TrES-4 is a slightly evolved, mildly 
metal-rich F-star with [Fe/H] $= 0.14\pm 0.09$. We determine stellar masses 
and radii using the combined spectro-photometric approach described in 
Sozzetti et al. (2007), in which $T_\mathrm{eff}$ and the normalized 
separation $a/R_\star$ are used in the comparison with stellar evolution models. 
We obtain $M_\star = 0.924_{-0.040}{+0.012}~M_\odot$, $R_\star =
0.813_{-0.027}^{+0.012}~R_\odot$ for TrES-3 and 
$M_\star = 1.384_{-0.046}{+0.070}~M_\odot$, $R_\star =
1.810_{-0.056}^{+0.071}~R_\odot$ for TrES-4. The improved planetary radius 
thus inferred confirms TrES-4 as the planet with the largest radius (and lowest 
density) discovered so far. Given the super-solar metallicity of its host, 
the TrES-4 system (together with WASP-1 and HAT-P-4) 
suggests that gas giant planet core mass is not a simple function of host-star 
metallicity or of radiation environment. 
We conclude that more definite statements on the relation of
the observations and planet structure theories can be made only
by reaching higher accuracy in the observed star/planet parameters.

\end{document}